\documentclass[prx,amsmath,amssymb,twocolumn,showpacs,superscriptaddress]{revtex4}

\usepackage{graphicx}
\usepackage{epstopdf}
\usepackage{dcolumn}
\usepackage{bm}
\usepackage{subfigure}
\usepackage{bm}
\usepackage{times}

\begin{document}
\title{Anisotropic Rabi model}
\author{Qiong-Tao Xie}
\affiliation{Beijing National Laboratory for Condensed Matter Physics,\\
and Institute of Physics, Chinese Academy of Sciences, Beijing
100190, China}
\affiliation{School of Physics and Electronic Engineering,
Hainan Normal University, Haikou 571158, China}

\author{Shuai Cui}
\affiliation{Beijing National Laboratory for Condensed Matter Physics,\\
and Institute of Physics, Chinese Academy of Sciences, Beijing
100190, China}
\author{Jun-Peng Cao}
\affiliation{Beijing National Laboratory for Condensed Matter Physics,\\
and Institute of Physics, Chinese Academy of Sciences, Beijing
100190, China}
\affiliation{Collaborative Innovation Center of Quantum Matter, Beijing 100190, China}

\author{Luigi Amico}
\email{lamico@dmfci.unict.it}
\affiliation{CNR-MATIS-IMM \& Dipartimento di Fisica e Astronomia,
Universit\'a Catania, Via S. Soa 64, 95127 Catania, Italy}
\affiliation{Center for
Quantum Technologies, National University of Singapore, 3 Science
Drive 2, 117543 Singapore}
\affiliation{Institute of Advanced Studies, Nanyang Technological
University, 1 Nanyang Walk, 637616 Singapore}

\author{Heng Fan}
\email{hfan@iphy.ac.cn}
\affiliation{Beijing National Laboratory for Condensed Matter Physics,\\
and Institute of Physics, Chinese Academy of Sciences, Beijing
100190, China}
\affiliation{Collaborative Innovation Center of Quantum Matter, Beijing 100190, China}

\date{\today}

\begin{abstract}
We define the anisotropic Rabi model as the generalization of the spin-boson Rabi model:
The Hamiltonian system breaks the parity symmetry; the rotating and counter-rotating interactions  are governed by two
different coupling constants; a further  parameter introduces a phase factor in the counter-rotating terms.
The exact energy spectrum  and eigenstates of the generalized model is worked out.
The solution is obtained as an elaboration of a recent proposed method for the isotropic limit of the model.
In this way, we  provide a long sought
solution  of a cascade of  models with immediate relevance in
different physical  fields, including {\it i)} quantum optics: two-level atom
in single mode cross electric and magnetic fields; {\it ii)} solid
state physics: electrons in semiconductors with Rashba and
Dresselhaus spin-orbit coupling; {\it iii)}  mesoscopic physics:
Josephson junctions flux-qubit quantum circuits.
\end{abstract}
\maketitle

\section{Introduction}

There are very simple settings in physics whose understanding
has very far reaching implications. This is the case of the Rabi
type models, that are  possibly the simplest 'organisms' describing
the interaction between a spin-half degree of freedom with a single boson.
Originally formulated in quantum optics to describe the atom-field interaction
\cite{Rabi}, such kind of models play a crucial role in many other
fields, especially with the advent of the quantum technologies.
Here, we introduce an anisotropic
generalization of the  Rabi model and discuss the exact energies  and eigenstates of it.
In this way, we provide a long sought solution  of a cascade of  models with
immediate relevance in various fields.

The Rabi type models provide the paradigm for key applications in a
variety of  different physical contexts, including quantum optics
\cite{Scully}, solid state and mesoscopic physics \cite{Wagner}.
Despite its importance, such models remained  intractable with exact
means for many years. Nevertheless, the physical community could
thoroughly analyze the Rabi model physics, essentially because the
physical settings allowed to easily adjust the field frequency  to
be  resonating with the atomic band-width. In this way,  assuming as
well that the field intensity is weak,  the Rabi model could be
drastically simplified  to the Jaynes-Cummings (JC) model \cite{JC}, through
the celebrated 'rotating wave'  approximation. The situation
radically changed in the last decade, when Quantum Technology has
been advancing towards more and more realistic applications
\cite{Raimond,Liebfried,Englund}.
In most  of the cases, if not all, the rotating wave approximation
cannot be applied. In the solid state applications, for example, the
electric field is an intrinsic quantity, that cannot be adjusted. On
the other hand, in the applications in mesoscopic physics (like
superconducting or QED circuits),  the most interesting regimes
correspond to very strong coupling between the spin variable and the
bosonic degree of freedom.

The class of the anisotropic Rabi model we consider in the present paper are described by the following Hamiltonian
\begin{eqnarray}
&&H =\omega a^{\dagger}a+\epsilon \sigma _x  +\Delta \sigma_z+ g (H_{r} +\lambda H_{cr}), \nonumber \\
&&H_r=(a^{\dagger}\sigma^{-}+a\sigma ^{+})\; , \nonumber \\
&&H_{cr}=e^{i\theta }a^{\dagger}\sigma^{+}+e^{-i\theta }a\sigma^{-} \;
\label{ha}
\end{eqnarray}
Here  $a^{\dagger}$ and $a$ are the creation and annihilation
operators for a bosonic mode of frequency $\omega$,
$\sigma^{\pm}=(\sigma _x\pm i\sigma _y)/2$, $\sigma _{x,y,z}$ are Pauli matrices for a two-level system, $2\Delta$
is  the energy difference between the two levels, $g$ denotes the
coupling strength of the rotating wave interaction
$a^{\dagger}\sigma^{-}+a\sigma^{+}$ between the two-level system and
the bosonic mode.
For simplicity, we already take the unit of $\hbar=1$.
In the Hamiltonian (\ref{ha}), the relative weight
between rotating and counter-rotating terms, denoted respectively as
$H_c$ and $H_{cr}$, can be adjusted by
tuning  the parameter $\lambda$. When $\epsilon=0$, the Hamiltonian
enjoys a discrete $Z_2$ symmetry
meaning that the parity of bosonic and spin excitations is
conserved.

Several attempts of solving these type of models were tried
employing Bethe ansatz and Quantum
Inverse Scattering techniques \cite{Irish,Amico-Frahm}. The isotropic
Rabi model corresponding to $\theta =0$ and $\lambda=1$ was solved
exactly in a seminal paper by Braak \cite{Braak}. Such an achievement
has allowed to explore the physics of the Rabi model in full
generality.

In this article, we present the exact solution of the
anisotropic Rabi models (\ref{ha}). We discuss how the models can
be applied to important physical settings in quantum optics,
mesoscopic and solid state physics. We also observe that such model
can be realized with cold atoms with arbitrary spin-orbit couplings.

\section{Exact analysis of the spectral problem}
To focus on the main results, we first provide a schematic of the  exact solution  of the spectral problem
\begin{equation}
H |\Psi\rangle=E |\Psi \rangle \; ,
\end{equation}
while
leaving the details in the appendix. 

Our approach elaborates on the method originally developed by Braak \cite{Braak}.
In order to find a concise solution, we
perform a unitary transformation on the spin degree of freedom in the Hamiltonian
$(\ref{ha})$. The eigenvalues can be found as,
\begin{eqnarray}
E_n=x_n-\frac {\lambda g^2}{\omega }. \label{eigenvalues}
\end{eqnarray}
where $x_n$ include regular and exceptional solutions.
The regular solutions are solely zeros of the transcendental function, while
the exceptional solutions are both
the zeros and poles leading to a finite transcendental function and energy degeneracy.
The transcendental function is as follows,
\begin{equation}
G_\epsilon (x)= \phi_1\overline{\phi}_2-\phi_2\overline{\phi}_1
\label{transcen}
\end{equation}
where
$
\phi_1(z)= \exp(-\frac{\sqrt{\lambda}
g\xi}{\omega}z)\sum_{n=0}^\infty
L_n^+(z+\frac{\sqrt{\lambda}g\xi^*}{\omega})^n$,
$\phi_2(z)= \exp(-\frac{\sqrt{\lambda}
g\xi}{\omega}z)\sum_{n=0}^\infty K_n^+
(z+\frac{\sqrt{\lambda}g\xi^*}{\omega})^n$, and
$\phi_1(-z)=\overline{\phi}_1(z)$,
$\phi_2(-z)=\overline{\phi}_2(z)$.
Fig. 1 and Fig. 2 displays the actual behavior of $G_\epsilon(x)$  in different parameter regimes.
For $\epsilon=0$, the $Z_2$ symmetry is recovered; in this case the transcendental function can be discussed through the functions $G_+=-e^{i\theta/2}\phi_1+\sqrt{\lambda} \phi_2$,  $G_-= e^{-i\theta/2}\phi_2+\sqrt{\lambda} \phi_1$, living in the two parity sectors, separately (see Fig. 2).
The explicit form of eigenfunctions $\phi _{1,2}(z)$ can also be obtained.

\begin{figure}[h]
\includegraphics[width=8cm]{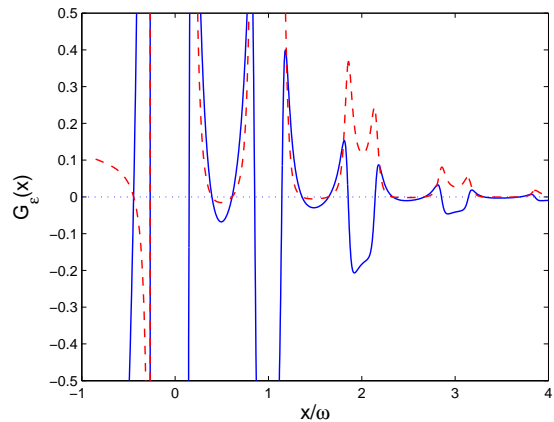}
\caption{(Color online)  Trascendental function $G_\epsilon(x)$  for $\epsilon\neq0$.
The parameters are $\omega=1$, $g=0.1$, $\lambda=0.5$, $\Delta=0.4$,
$\epsilon=0.2$, and $\theta=-\pi/2$, the zero points whose real
(blue-solid line) and imaginary part (red-dashed line)  of $G_\epsilon$ both equal
0 correspond the eigenvalues of Hamiltonian.} \label{sup-epsi02-main}
\end{figure}

\begin{figure}[tb]
\includegraphics[width=8cm]{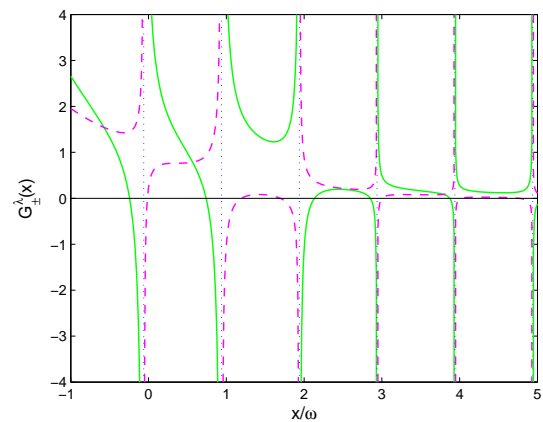}
\caption{(color online)  Trascendental functions
$G_{+}^{\lambda}(x)$ (green-solid line) and real
part of $G_{-}^{\lambda}(x)$ (purple-dashed line)   for $\epsilon=0$.  The parameters are
$\omega=1$, $g=0.7$, $\Delta=0.4$, $\lambda=0.5$ and
$\theta=-\pi/2$. The regular parts of the energy spectrum are
determined by the zeros of the transcendental function
$G_{\pm}^{\lambda}(x)$, the dotted vertical lines denote the
poles $x\approx n-0.0612$, $n=0,1,2,...$, see (\ref{addpole}). Notice that the imagine part of
$G_{-}^\lambda(x)$ gives the same zero and poles as the real part,
which is not shown. }\label{GFZ2-main}
\end{figure}

\begin{figure}
\includegraphics[width=8cm]{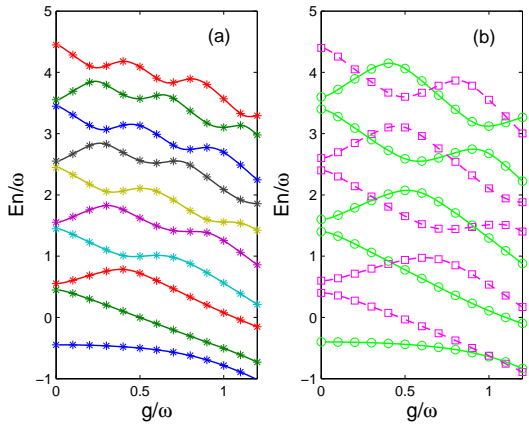}
\caption{(Color online) Comparison between exact solution and the
numerical results. (a) Energy spectrum for $\omega=1$, $\Delta=0.4$,
$\lambda=0.5$, $\epsilon=0.2$ and $\theta=-\pi/2$. Solid lines
are exact results, and the energy levels are differentiated by colors.
Numerical results are represented by stars. (b) Energy
spectrum for $\omega=1$, $\Delta=0.4$, $\lambda=0.5$, $\epsilon =0$,
and $\theta=0$ in the spaces with positive (green lines with circles) and negative
(purple lines with squares) parities. Small squares and circles represent numerical results
The first energy level crossing point is at
$g_c=4/\sqrt{15}\approx 1.0328$ and $E=-2/3$ which has no definite parity.}
\label{figure1}
\end{figure}

For vanishing $\epsilon $ or  multiple of $\omega /2$, the system enjoys a  $Z_2$ (parity) symmetry.
In this case,  the energy spectrum  can be labeled by the two eigenvalues of
the parity operator (corresponding to green-with-circle and
purple-with-square lines in Fig. 3(b)). At the points of level
crossings the energy is  doubly
degenerate. For the isotropic case, those solutions were  found previously by
Judd \cite{Judd,Braak}. For our anisotropic Rabi model,
the crossing points are found as
$E_n=n\omega-(1+\lambda^2)^2/2\omega $, corresponding to exceptional spectrum.
This exceptional spectrum is characterized by the merging of a pole
with a zero resulting in a finite, nonzero transcendental function in Eq. (\ref{transcen})
at energies corresponding to Juddian solutions \cite{Judd}, see also Section VI.

For non vanishing generic values of  $\epsilon $, the $Z_2$ symmetry is lost.
This is manifested in the spectrum; in particular,  there are no degeneracies (see  Fig. 3(a)).
As we shall see, the parameter $\theta $ is important to capture  general spin-orbit couplings.
We remark that with $Z_2$ symmetry
preserved, $\theta $ can be deleted by a unitary transformation,  and thus it does
not change the energy spectrum.
When $Z_2$ symmetry is broken with non-vanishing $\epsilon $,
the parameter $\theta $ enters the  energy spectrum through $\epsilon \sigma _x$
and this unitary transformation will induce term $\sigma _x\rightarrow \cos (\theta /2)\sigma _x
+\sin (\theta /2)\sigma _y$.

As it will be later argued to be important for many applications,
we  quantify on  the energy correction due to the counter-rotating term (Bloch-Siegert shift \cite{Bloch-Siegert}).
Based on the exact solution, we can give  closed expressions  in several  interesting limits.
For $2\Delta \approx \omega $, $g\ll \omega $ the shift is
$g^2/\omega $.
For $\epsilon=0$, at the degenerate points, and setting  $|\Delta| =(1-\lambda ^2)g^2/2\omega$, the ground state
energy gap between the JC model and the anisotropic Rabi model can
be found as,
\begin{eqnarray}
\Delta E_0=\frac {\lambda ^2g^2}{\omega }.
\end{eqnarray}
For  $\lambda =1$, it is just the standard Bloch-Siegert shift\cite{Bloch-Siegert,Shirley}.
Such gap can be obtained also for the excitations. For the first and the second excited states
at degenerate points, it reads  $\sim \lambda ^2g^4/\omega^3$.

\section{Applications}
In this section, we  discuss how our solution can contribute  to approach  important problems in different physical contexts.
Specifically, we will consider applications in  quantum optics,  mesoscopic physics, and spintronics.
\subsection{Application to quantum optics: two level atom in cross electric
and magnetic field.}  When an atom is subjected of a crossed
electric and magnetic field, the selection rules are not dictated by
the possible values of the atomic angular momentum. Therefore, both
the electric dipole and magnetic dipole transition are allowed. The
Hamiltonian describing the system is
\begin{equation}
H=H_0 - \mathbf{d}\cdot \mathbf{E}-\mathbf{\mu}\cdot \mathbf{B}
\label{quantum_optics}
\end{equation}
where we have assumed that the quadrupole transitions can be
neglected. Inserting the standard expressions of the  quantized
electric and magnetic fields are respectively $E\sim (a+a^\dagger)$
and  $B\sim i (a-a^\dagger)$ , Eq.(\ref{quantum_optics}) can be
recast into our anisotropic Rabi model Eq.(\ref{ha}) with
\begin{eqnarray}
g &=& {{\langle + | d|- \rangle + \langle + | \mu |- \rangle}\over{2}} \\
\lambda &= &{{\langle + | d|- \rangle - \langle + | \mu |-
\rangle}\over{\langle + | d|- \rangle + \langle + | \mu |- \rangle}}
\end{eqnarray}
being $H_0|\pm\rangle =E_{\pm}|\pm \rangle$.

\subsection{Application to superconducting circuits.}
\label{sc_circuit}
Superconducting
circuits exploits the inherent coherence of superconductors for a
variety of technological applications, including quantum
computation\cite{esteve} In this case, the bosonic fields typically
represent the electromagnetic fields generated by the
superconducting currents. The spin degree of freedom describes the
two states of the qubit.

As immediate application, we consider
 two inductively coupled
dc-Superconducting Quantum Interference Devices
(SQUIDs)\cite{CHIORESCU,MURALI}: a primary SQUID $p$ (assumed large
enough to produce an electromagnetic field characterized by a
bosonic mode) controls the qubit realized by the secondary SQUID. In
the limit of negligible capacitive coupling between the two SQUID's,
the circuit  Hamiltonian is
\begin{equation}
{\cal H}_{circuit}=\omega_p a^\dagger a - 2 E^s_{J}  \sigma^x
-2 \tilde{L}_p(a+a^\dagger) \sigma^x
- i M (a-a^\dagger) \sigma^y  
\label{H-circuit}
\end{equation}
where $\omega_p$  is the ``frequency''  of the primary  and
$E_J^s=E_J^s(\phi_{ext}) $ provides the level splitting of the   secondary
SQUID, controlled by the external magnetic field;  $\tilde{L}_p$ and $M$ are fixed by the inductance of the circuit and  the mutual
inductance  respectively; the  gate voltage  $V_g$  is tuned to the charge degeneracy point. 
The Eq.(\ref{H-circuit}) can be recast into the anisotropic Rabi
model\cite{AmicoHikami}: $\displaystyle{\{\omega_p,E^s_J, 2
\tilde{L}_p, M \} \rightarrow \{\omega, \epsilon , g(1+\lambda),
g(1-\lambda)\}}$.

\begin{figure}
\includegraphics[width=8cm]{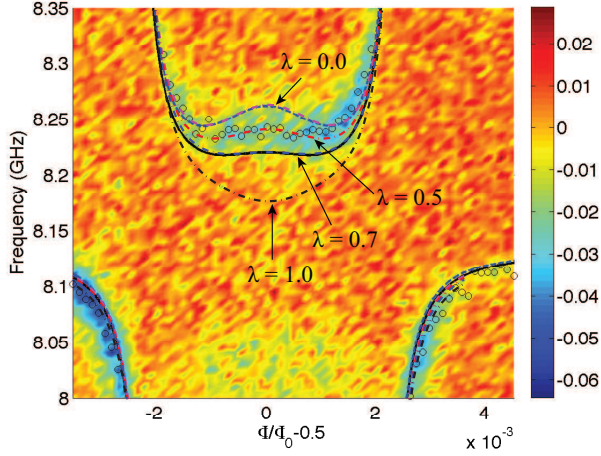}
\caption{(Color online) Comparison between the anisotropic Rabi model with
experimental results.
The spectra (dashed dot lines) of the anisotropic Rabi
model with $\lambda =0.0, 0.5, 0.7, 1.0$, the other parameters are
$g=0.74$GHz, $\Delta =4.21$GHz, $I_p=500$nA, $\omega _r=8.13$GHz,
the same as previous investigated \cite{Mooij}. The curves with $\lambda =0.5$ (red dashed
dot line) agree perfectly with the experimental data (circle points).}
\label{datafitting}
\end{figure}

We comment that the implications of the simultaneous presence of the
rotating and counter-rotating terms have been evidenced
experimentally \cite{Mooij,Niemczyk,Hakonen}. The experimental system is an
LC resonator magnetically coupled to a superconducting flux qubit in
the ultrastrong coupling regime. Indeed the experimental data were
interpreted as   Bloch-Siegert energy correction of the
Jaynes-Cummings dynamics. Here we point out that the experimental
results can be fitted very well in terms of our  anisotropic
Rabi model, see Fig.\ref{datafitting} (further details are provided in the appendix \ref{details_superconducting}).
This provides an indication that, the inductance of the circuit is, indeed, very different from the mutual inductance between the primary and the qubit.


\subsection{Applications to electrons in semiconductors with spin-orbit
coupling.} Spin-orbit coupling effects have been opening up new
perspectives in solid state physics, both for fundamental research
(including topological insulators and spin-Hall
effects\cite{topo_ins,spinHall}) and applications (notably
spintronics\cite{Datta}). Electronic spin orbit coupling can be
induced by the electric field acting at the two dimensional
interfaces of semiconducting heterostructure
devices\cite{Datta,Rashba,Dresselhaus,Molenkamp,Loss}. The effective Hamiltonian
reads
\begin{eqnarray}
H&=&{{1}\over{2m}} \pi^2+{{1}\over{2}}  g \mu_B B \sigma_z +H_{so} \nonumber \\
H_{so}&=& H_R+H_D \nonumber \\
H_R&=&\alpha (\pi_x \sigma_y-\pi_y \sigma_x) \; , \; H_D=\beta
(\pi_x \sigma_x-\pi_y \sigma_y) \label{application_spinorbit}
\end{eqnarray}
where $\mathbf{\pi}=\{\pi_x,\pi_y,\pi_z\} $ is the electrons
canonical momentum $\mathbf{\pi}=\left (\mathbf{p}
-{{q}\over{c}}\mathbf{A} \right )$. $H_R$ and $H_D$ are the Rashba \cite{Rashba}
and Dresselhaus \cite{Dresselhaus} spin-orbit interactions. The coupling constant
$\alpha$  depends on the  electric field across the well, while the
Dresselhaus coupling $\beta$ is determined by the geometry of the
hetereostructure. The perpendicular magnetic field couples both to
the electronic spin and orbital angular momentum. Applying the
standard procedure leading to the Landau levels, the Hamiltonian
(\ref{application_spinorbit}) can be recast into our anisotropic
Rabi model:
$\alpha=  \sqrt{g^2+(1+\lambda)^2} \sin{\theta}$,
$\beta=  \sqrt{g^2+(1+\lambda)^2} \cos{\theta}.$
Incidentally, we observe that the simultaneous presence of
Dresselhaus and Rashba contributions couples all the Landau levels,
making our exact solution immediately relevant for the physics of
the system.

We comment that the Hamiltonian (\ref{application_spinorbit}), has
been realized with cold fermionic atoms systems, opening the avenue
to study the spin-orbit effects with controllable parameters and in
extremely clean environments \cite{Lin,Liu,spinorbitketterle,Bloch}.

\section{Entanglement entropy}
In this section we elaborate on the phenomenon displayed in the Fig.\ref{figure1}: For the anisotropic Rabi model, level crossings occur between  eigenvalues of different parity sectors.

The crossing between the ground state and the first excited state occurs for the anisotropic case which
corresponds to the exact solutions obtained by Judd \cite{Judd}, see also appendix B and FIG.{\ref{Figure5}}. This
does not occur in the isotropic Rabi model,
and is possibly due to the competition between the rotating and counter-rotating interaction terms.
The position of this point can be
analytically determined by the relation $K_1(x_0^{pole})=0$ as
mentioned above, i.e., $a_0=0$, $b_0=0$,
\begin{eqnarray}
g&=& \sqrt{\frac{2|\Delta|\omega}{1-\lambda^2}},
\label{specialg}
 \\
E_0&=& -\frac{(1+\lambda^2)g^2}{2\omega}.
\label{speciale}
\end{eqnarray}
For the crossing of
the ground state and the first excited state, we find that the series terminates
at the first term, as $K_0=1$ and
$L_0=2\sqrt{\lambda}\xi^*/(1-\lambda)$, thereby resulting in
\begin{eqnarray}
\phi _1&=&\frac{2\sqrt {\lambda }\xi^*}{1-\lambda }\exp(-\frac {\sqrt {\lambda }g\xi}{\omega }z),
\label{solution1} \\
\phi _2&=&\exp(-\frac {\sqrt {\lambda }g\xi}{\omega }z) \label{solution2}.
\end{eqnarray}

\begin{figure}
\includegraphics[width=8cm]{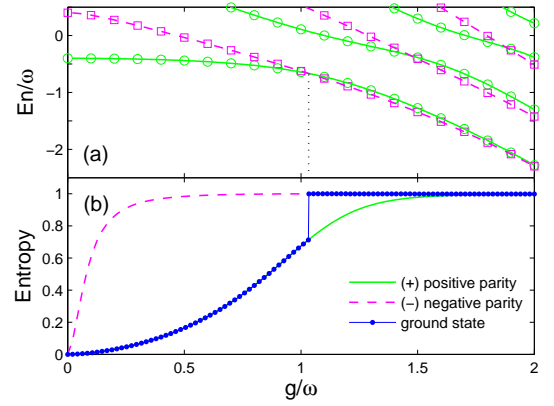}
\caption{(color online) Energies (upper panel) and entanglement entropies (lower panel)
of the ground state and the first excited state.
A level crossing occurs at $g_c=4/\sqrt {15}$, marking a change of the parity of the ground state:
$g<g_c$ the parity is positive (green-solid), and negative (purple-dashed) for $g>g_c$,
as shown in upper panel.  Correspondingly,
we find that  the ground state entropy
displays a sharp discontinuity  at the level crossing point, as shown in lower panel.
}
\label{Figure7}
\end{figure}

Here, we study the entanglement entropy of the ground state.
It can be obtained by calculating the von Neumann entropy of the spin state
by tracing out the bosonic degree of freedom from the eigenstates (\ref{parity1+},\ref{parity1-}),
see \cite{Amico-RMP,Cramer-RMP} for methods.
We observe that the level crossing and change of symmetry of the ground state are reflected in a clear discontinuity of the entanglement entropy  (see Fig.\ref{Figure7}).
We remark that the ground states are degenerate at the level crossing point which
is special.

\section{Discussion}

In this article, we discussed a carefully chosen generalization of the Rabi model: The Hamiltonian system breaks the parity symmetry; the rotating and counter-rotating interactions  are governed by two
different coupling constants; a further  parameter introduces a phase factor in the counter-rotating terms.
We obtained exact energies and eigenstates of the system through the analytical properties of a transcendental function.
We note that, {\it because of the  anisotropic coupling}  a peculiar phenomenon occurs in the energy spectrum of the system: the eigenstates belonging to different parity sectors swap in couples. We have quantified the crossing between the ground and the first excited state through the entanglement entropy of the spin system.

Our Hamiltonian systems capture the physics of  notoriously important problems in different physical contexts, including  two dimensional electron gas with general spin-orbit interaction, two level atom  in electromagnetic field,  and superconducting circuits in ultra-strong regimes.
We explained how our results are immediately relevant for the experimental situations.

We believe that  superconducting circuits made of two coupled SQUID's could provide access   to a systematic  experimental study of  the physical effects of the anisotropic  Rabi interaction. Specifically, our study indicates that the circuit inductance, SQUID-SQUID inductance and the external magnetic field  are the parameters that should be varied  to study the crossover from the weak to strong coupling regimes (see Sect.\ref{sc_circuit}).

\emph{Acknowledgements.} Q. T. X. and S. C. contributed equally. We thank useful discussions with B. Englert and Wei-Bin
Yan. We thank X. B. Zhu for discussions about the experimental realization in superconducting flux qubit system.
This work was supported by ``973'' program (2010CB922904),
grants from NSFC and CAS.

\begin{appendix}

\section{Exact solution of the anisotropic Rabi model}
For the Hamiltonian presented in (\ref{ha}),
the parameter $\lambda$ controls the anisotropy between the rotating
and the counter rotating terms and $\theta$ introduce a phase into the counter rotating terms only;
term $\epsilon\sigma_x$ breaks the $Z_2$ symmetry, and therefore the eigenspace of the
model (\ref{ha}) cannot be split in  invariant subspaces. Nevertheless, the model  (\ref{ha}) can still be solved exactly with the approach originally developed by Braak\cite{Braak} for the isotropic model $\lambda=1$, $\theta=\epsilon=0$.

In solving exactly this model, firstly, for technical convenience (we comment further below),
we perform a unitary transformation $U(\lambda, \theta)$,
\begin{eqnarray}
U(\lambda,\theta)&=&\left(\begin{array}{cc} \cos \eta e^{i\theta/2} & -\sin \eta \\
\sin\eta & \cos \eta e^{-i\theta/2}
\end{array}\right)
\nonumber \\
&=&\frac{1}{\sqrt{1+\lambda}} \left(\begin{array}{cc} \xi & -\sqrt{\lambda} \\
\sqrt{\lambda} & \xi^*
\end{array}\right),
\label{transformationU}
\end{eqnarray}
where $\tan \eta=\sqrt{\lambda}$, $\xi=e^{i\theta/2}$, when
$\theta=0$, $\xi=1$, it is a orthogonal transformation.
The Hamiltonian (\ref{ha}) becomes
\begin{widetext}
\begin{equation}
U^\dagger H U=\left(\begin{array}{cc} \omega a^\dagger
a+\sqrt{\lambda}g(\xi^* a+\xi a^\dagger)+c &
\xi^{*2}(1-\lambda)g a -d^* \\
\xi^2 (1-\lambda)g a^\dagger -d & \omega a^\dagger
a-\sqrt{\lambda}g(\xi^* a+ \xi a^\dagger)- c
\end{array} \right).
\label{Hamiltonian1}
\end{equation}
where
$c=\frac{1-\lambda}{1+\lambda}\Delta+\frac{(\xi+\xi^*)\sqrt{\lambda}}{1+\lambda}\epsilon$,
$d=\frac{2\sqrt{\lambda}\xi}{1+\lambda}\Delta-\frac{\xi^2-\lambda}{1+\lambda}\epsilon$.
We exploit the Bargmann
representation of bosonic operators in terms of
analytic functions:  $a^{\dagger }\rightarrow z,
a\rightarrow \frac {\partial }{\partial z}$, and consider the
eigenfunction of the Hamiltonian as $(\phi _1, \phi _2)^T$, we have
\begin{eqnarray}
\left[\omega z\frac{{\rm d}}{{\rm d}z}+\sqrt{\lambda}g (\xi^*
\frac{{\rm d}}{{\rm d}z}+\xi z)+c\right]\phi_1+
\left[\xi^{*2}(1-\lambda)g\frac{{\rm d}}{{\rm d}z}-d^*
\right]\phi_2&=&
E\phi_1, \label{Barg1}\\
\left[\xi^2 (1-\lambda)g z-d \right]\phi_1+\left[ \omega z\frac{{\rm
d}}{{\rm d}z}-\sqrt{\lambda}g(\xi^* \frac{{\rm d}}{{\rm d}z}+\xi
z)-c\right]\phi_2&=&E\phi_2. \label{Barg2}
\end{eqnarray}
\end{widetext}
For convenience, we introduce the notations
$\phi_{1,2}(z)=\exp(-\frac{\sqrt{\lambda}g\xi}{\omega}z)\psi_{1,2}(y)$,
$y=z+\frac{\sqrt{\lambda}g\xi^*}{\omega}$, $x=E+\frac{\lambda
g^2}{\omega}$, $f=d+\frac{(1-\lambda)\sqrt{\lambda}g^2
\xi}{\omega}$.
Now, we obtain,
\begin{eqnarray}
(\omega y\frac{{\rm d}}{{\rm d}
y}-x+c)\psi_1 = \left[f^*-\xi^{*2}(1-\lambda)g\frac{{\rm d}}{{\rm d} y}\right]\psi_2, \label{sup-psi1}
\end{eqnarray}
and
\begin{eqnarray}
&&\left[(\omega y-2\sqrt{\lambda}g\xi^*)\frac{{\rm d}}{{\rm
d}y}-2\sqrt{\lambda}g \xi y+\frac{4\lambda g^2}{\omega}-x-c
\right]\psi_2 \nonumber \\
&&~~~~~~~~~~~~~~~
= \left[f-(1-\lambda)g y\right]\psi_1.
\label{sup-psi2}
\end{eqnarray}
Assuming that the  functions $\psi_{1,2}$ can be expanded as,
$\psi_2=\sum_{n=0}^\infty K_n^+(x)y^n$, $\psi_1=\sum_{n=0}^\infty
L_n^+(x)y^n$, from Eq. (\ref{sup-psi1}), the relation between
$K_n^+$ and $L_n^+$ is found as
\begin{eqnarray}
L_n^+ =\frac{f^* K_n^+ -\xi^{*2}(1-\lambda)g K_{n+1}^+(n+1)}{n\omega
-x+c}.
\end{eqnarray}
Then from Eq.(\ref{sup-psi2}), the recursive relation of $K_n^+$ is
obtained,
\begin{eqnarray}
a_n(x)K_{n+1}^+ &=& b_n(x)K_{n}^+ +c_n(x)K_{n-1}^+, \label{sup-an1}\\
a_n(x)&=& \left[2\sqrt{\lambda}-\frac{
(1-\lambda)f \xi^*}{n\omega-x+c}\right](n+1)g\xi^* ,\label{sup-an} \\
b_n(x)&=& \frac{4\lambda g^2}{\omega}+n\omega-x-c- \frac{ f^*
f}{n\omega-x+c} \nonumber \\
&&
-\frac{(1-\lambda)^2
g^2 n}{(n-1)\omega-x+c}, \label{sup-bn} \\
c_n(x) &=& -2\sqrt{\lambda}g\xi+\frac{(1-\lambda)gf^*
\xi^2}{(n-1)\omega-x+c}. \label{sup-cn}
\end{eqnarray}
where $K_{-1}^+=0, K_0^+=1$, $n=0,1,2,...$. Incidentally, we comment that the unitary transformation (\ref{transformationU})
is a key step leading to simplify  the recursive relations which
involve only three terms, as it is displayed above.

Consequently, one sets of solutions is  obtained:
\begin{eqnarray}
\phi_1(z)&=& \exp(-\frac{\sqrt{\lambda}
g\xi}{\omega}z)\sum_{n=0}^\infty
L_n^+(z+\frac{\sqrt{\lambda}g\xi^*}{\omega})^n, \label{phi1z} \\
\phi_2(z)&=& \exp(-\frac{\sqrt{\lambda}
g\xi}{\omega}z)\sum_{n=0}^\infty K_n^+
(z+\frac{\sqrt{\lambda}g\xi^*}{\omega})^n. \label{phi2z}
\end{eqnarray}

Then, substituting
$z\rightarrow -z$ in Eq.(\ref{Barg1},\ref{Barg2}),  $\phi_1(-z)=\overline{\phi}_1(z)$,
$\phi_2(-z)=\overline{\phi}_2(z)$ are eigenfunctions of the spectral problem (\ref{Barg2}) as well.
Such functions can be obtained by applying the same procedure led to  (\ref{phi1z}) and  (\ref{phi2z}).
The
differential equations for $\overline{\phi}_1(z)$ and  $\overline{\phi}_2(z)$ are
\begin{eqnarray}
&&\left[\omega z\frac{{\rm d}}{{\rm d}z}-\sqrt{\lambda}g (\xi^*
\frac{{\rm d}}{{\rm d}z}+\xi
z)+c\right]\overline{\phi}_1+\nonumber \\
&&+
\left[-\xi^{*2}(1-\lambda)g\frac{{\rm
d}}{{\rm d}z}-d^*\right]\overline{\phi}_2=
E\overline{\phi}_1,\\
&&\left[-\xi^2 (1-\lambda)g z-d\right]\overline{\phi}_1+
\nonumber \\
&&+\left[ \omega
z\frac{{\rm d}}{{\rm d}z}+\sqrt{\lambda}g(\xi^* \frac{{\rm d}}{{\rm
d}z}+\xi z)-c\right]\overline{\phi}_2=E\overline{\phi}_2.
\end{eqnarray}
Using the following notations,
$\overline{\phi}_{1,2}(z)=\exp(-\frac{\sqrt{\lambda}g\xi}{\omega}z)\overline{\psi}_{1,2}(y)$,
$y=z+\frac{\sqrt{\lambda}g\xi^*}{\omega}$, $x=E+\frac{\lambda
g^2}{\omega}$, $\overline{f} =d-\frac{(1-\lambda)\sqrt{\lambda}g^2
\xi}{\omega}$, the above equations can be rewritten as,
\begin{eqnarray}
\left[(\omega y-2\sqrt{\lambda}g\xi^*)\frac{{\rm d}}{{\rm
d}y}-2\sqrt{\lambda}g\xi y+\frac{4\lambda g^2}{\omega}
-x+c\right]\overline{\psi}_1 \nonumber \\
=
\left[\overline{f}^*+\xi^{*2}(1-\lambda)g \frac{{\rm
d}}{{\rm d} y}\right]\overline{\psi}_2, \label{sup-psi22} \\
(\omega y\frac{{\rm d}}{{\rm d} y}-x-c)\overline{\psi}_2 =
\left[\overline{f}+\xi^2(1-\lambda)g y\right]\overline{\psi}_1.
\label{sup-psi21}
\end{eqnarray}
Expand functions $\overline{\psi}_{1,2}$ as
$\overline{\psi}_1=\sum_{n=0}^\infty K_n^-(x)y^n$,
$\overline{\psi}_2=\sum_{n=0}^\infty L_n^-(x)y^n$, from
Eq.(\ref{sup-psi21}) we find the relation of $K_n^-$ and $L_n^-$,
\begin{eqnarray}
L_n^- = \frac{\overline{f} K_n^- +\xi^2 (1-\lambda)g
K_{n-1}^-}{n\omega -x-c}
\end{eqnarray}
Then from Eq.(\ref{sup-psi22}) we obtain the recursive relation
\begin{eqnarray}
\overline{a}_n(x)K_{n+1}^- &=& \overline{b}_n(x)K_{n}^- +\overline{c}_n(x)K_{n-1}^-, \label{sup-an2} \\
\overline{a}_n(x)&=& \left[2\sqrt{\lambda}+\frac{
(1-\lambda)\overline{f}\xi^{*} }{(n+1)\omega-x-c}\right](n+1)g\xi^* ,  \nonumber \\
\\
\overline{b}_n(x)&=& \frac{4\lambda g^2}{\omega}+ n\omega -x +c-
\frac{ \overline{f}^* \overline{f}}{n\omega-x-c}
\nonumber \\
&&-\frac{(1-\lambda)^2
g^2 (n+1)}{(n+1)\omega-x-c}, \\
\overline{c}_n(x) &=&
-2\sqrt{\lambda}g\xi-\frac{(1-\lambda)g\overline{f}^*
\xi^2}{n\omega-x-c}.
\end{eqnarray}
where $K_{-1}^-=0, K_0^-=1$, $n=0,1,2,...$

Going back to the original notations, we have,
\begin{eqnarray}
\overline{\phi}_1(z)&=& \exp(-\frac{\sqrt{\lambda}
g\xi}{\omega}z)\sum_{n=0}^\infty
K_n^-(z+\frac{\sqrt{\lambda}g\xi^*}{\omega})^n, \label{phibar1}\\
\overline{\phi}_2(z)&=& \exp(-\frac{\sqrt{\lambda}
g\xi}{\omega}z)\sum_{n=0}^\infty L_n^-
(z+\frac{\sqrt{\lambda}g\xi^*}{\omega})^n. \label{phibar2}
\end{eqnarray}

Considering the relation of these two sets of eigenstates
mentioned above, $\phi_1(-z)=C\overline{\phi}_1(z)$,
$\phi_2(-z)=C\overline{\phi}_2(z)$, then canceling the arbitrary
constant $C$, a transcendental function can be constructed as,
\begin{eqnarray}\label{G_epsilon}
G_{\epsilon }(x;
z)&=&\phi_1\overline{\phi}_2-\phi_2\overline{\phi}_1,
\end{eqnarray}
Because $G_{\epsilon}(x;z)$ is well defined at $z=\pm
\sqrt{\lambda}g\xi^*/\omega$ within the convergent radius
$R=2\sqrt{\lambda}g\xi^*/\omega$, we can set $z=0$
\cite{Braak}. The function $G_\epsilon(x; 0)$ is analytic in the complex plane except
in the simple poles
\begin{eqnarray}
x^{pole}_n &=&n\omega-\frac{(1-\lambda)^2
g^2}{2\omega}+\frac{(\xi+\lambda\xi^*)\epsilon}{2\sqrt{\lambda}}, \\
\overline{x}_n^{pole} &=&(n+1)\omega-\frac{(1-\lambda)^2
g^2}{2\omega}-\frac{(\xi+\lambda\xi^*)\epsilon}{2\sqrt{\lambda}},
\end{eqnarray}
which follows from  the zeros of the denominator of $K_n^{\pm}$:
$a_n(x)=0$ and $\overline{a}_n(x)=0$ in Eq.(\ref{sup-an1}) and
Eq.(\ref{sup-an2}), respectively. Then, the eigenvalues and
eigenstates can be obtained by solving $G_\epsilon(x)=0$,
\begin{eqnarray}
E_n&=&x_n-\frac{\lambda g^2}{\omega}, \label{eigenvalue} \\
\Psi_n&=& U(\lambda,\theta)\left(\begin{array}{cc} \phi_1(x_n) \\
\phi_2(x_n)
\end{array} \right)
\nonumber \\
&=& U(\lambda,\theta)
\left(\begin{array}{cc} \sum_{n=0}^\infty L_n^+ |n\rangle\rangle \\
\sum_{n=0}^\infty K_n^+ |n\rangle\rangle \end{array} \right),
\label{eigenstate}
\end{eqnarray}
Using the second solution,  the eigenstates of
Hamiltonian with  $a\rightarrow -a$, $a^\dagger \rightarrow
-a^\dagger$ can be obtained:
\begin{eqnarray}
\overline{\Psi}_n&=& U(\lambda,\theta)\left(\begin{array}{cc} \overline{\phi}_1(x_n) \\
\overline{\phi}_2(x_n)
\end{array} \right)
\nonumber \\
&=& U(\lambda,\theta)
\left(\begin{array}{cc} \sum_{n=0}^\infty K_n^- |n\rangle\rangle \\
\sum_{n=0}^\infty L_n^- |n\rangle\rangle \end{array} \right),
\end{eqnarray}
where
\begin{eqnarray}
&&|n\rangle\rangle \doteq (a^\dagger+\frac{\sqrt{\lambda}g\xi^*}{\omega})^n|-\frac{\sqrt{\lambda}g\xi}{\omega}\rangle,
\nonumber \\
&&|-\frac{\sqrt{\lambda}g\xi}{\omega}\rangle =e^{-\frac{\lambda
g^2}{2\omega^2}-\frac{\sqrt{\lambda}g\xi}{\omega}
a^\dagger}|0\rangle.
\end{eqnarray}

In case $\lambda=1$ and $\theta=0$, we can recover the results given by
Braak \cite{Braak} and some generalized results \cite{Chen,Tomka,Albert2012}.
Some other detailed calculations can be found in appendix.

\section{results for $Z_2$ symmetric case}

For $\epsilon =0$, the anisotropic Rabi model
enjoys a  $Z_2$ symmetry reflecting the conservation of the parity of the operator
\begin{eqnarray}
\hat{N}=a^\dagger a +\frac{1}{2}(\sigma_z+1),
\end{eqnarray}
In this case, the  phase factors
$e^{\pm i \theta}$  in the Hamiltonian can be canceled by a unitary
transformation $R(\theta)$,
\begin{equation}
R(\theta)=e^{i\frac{\theta}{2}(\hat{N}-\frac{1}{2})}=e^{i\frac{\theta}{2}(\frac{\sigma_z}{2}+a^\dagger
a)},
\end{equation}
\begin{eqnarray}
R^\dagger(\theta) H R(\theta)&=&\omega a^\dagger a +\Delta \sigma_z
+g[\sigma^+ a+\sigma^- a^\dagger
\nonumber \\
&&+\lambda (\sigma^+ a^\dagger
+\sigma^- a)].
\label{ha-z}
\end{eqnarray}
Therefore the parameter $\theta$ gives no contribution to the energy spectra, but  enters the wave functions only.

We shall see that the  $Z_2$ symmetry effectively simplifies  the procedure of finding the exact spectrum since the transcendental function
$G_\epsilon(x)$, $\epsilon =0$,  can be discussed into the different parity sectors separately.

To simplify the solution of the spectral problem, we resort to a similar trick we employed above. Namely, we apply  the rotation
\begin{eqnarray}
V&=&U^{\dagger }(\lambda, \theta )W
\nonumber \\
&=&\frac{1}{\sqrt{2(1+\lambda)}} \left(\begin{array}{cc} \xi^*+\sqrt{\lambda} & -\xi^*+\sqrt{\lambda} \\
\xi-\sqrt{\lambda} & \xi+\sqrt{\lambda}
\end{array} \right).
\end{eqnarray}
with
\begin{equation}
W=\frac{1}{\sqrt{2}}\left(\begin{array}{cc} 1 & -1 \\
1 & 1
\end{array} \right).
\label{W-transform}
\end{equation}
to the Hamiltonian (\ref{ha-z}). Now, we have the Hamiltonian,
\begin{widetext}
\begin{eqnarray}
W^{\dagger }HW=
\left( \begin{array}{cc}
\omega a^{\dagger }a+\frac {1+\lambda }{2}(a+a^{\dagger })
& \frac {1-\lambda }{2}(a-a^{\dagger })+\Delta \\
-\frac {1-\lambda }{2}(a-a^{\dagger })+\Delta &
\omega a^{\dagger }a-\frac {1+\lambda }{2}(a+a^{\dagger })
\end{array} \right) .
\end{eqnarray}
\end{widetext}

\begin{figure}
\includegraphics[width=8cm]{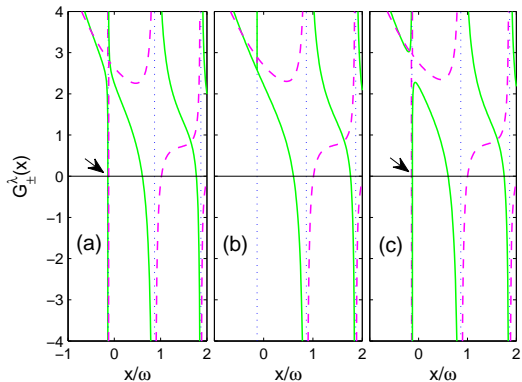}
\caption{(color online) Comparison of transcendental functions
near ground state degenerate point. From left to right, the parameters are (a) $g=g_c-0.01$,
(b) $g=g_c$, (c) $g=g_c+0.01$, where $g_c=4/\sqrt {15}$ which is the degenerate point.
 $G_{+}^{\lambda}(x)$ (green-solid line) and
$G_{-}^{\lambda}(x)$ (purple-dashed line) are presented as functions of $x$,
where $\omega=1$, $\Delta=0.4$, $\lambda=0.5$,
$\theta=0$. The difference of (a,b,c) can be observed, for
example, at the points marked by arrows in the figures.
For (a) at the point marked by arrow,
the energy of positive parity is slightly less than the energy
of negative parity, it is reversed in (c), while in (b),
the pole is lifted since $K_1=0$.}\label{Figure5}
\end{figure}

The eigenfunctions $\phi _1,\phi _2$ (and similarly  $\overline{\phi}_1, \overline{\phi}_2$)
in the main text transform according to  $(\varphi_1,
\varphi_2)^T=V^\dagger (\phi_1, \phi_2)^T$:
\begin{eqnarray}
\varphi_1&=&\frac{(\xi+\sqrt{\lambda})\phi_1+(\xi^*-\sqrt{\lambda})\phi_2}{\sqrt{2(1+\lambda)}},
\\
\varphi_2&=&\frac{(-\xi+\sqrt{\lambda})\phi_1+(\xi^*+\sqrt{\lambda})\phi_2}{\sqrt{2(1+\lambda)}}.
\end{eqnarray}
We know that $\phi _1,\phi _2$ read as,
\begin{eqnarray}
\phi_1(z)&=& \exp(-\frac{\sqrt{\lambda
}g\xi}{\omega}z)\sum_{n=0}^{\infty}L_n(x)(z+\frac{\sqrt{\lambda}g\xi^*}{\omega})^n, \nonumber \\
\phi_2(z)&=& \exp(-\frac{\sqrt{\lambda
}g\xi}{\omega}z)\sum_{n=0}^{\infty}K_n(x)(z+\frac{\sqrt{\lambda}g\xi^*}{\omega})^n,
\nonumber \\
\end{eqnarray}
where
\begin{eqnarray}
L_n &=&\frac{f^* K_n -\xi^{*2}(1-\lambda)g K_{n+1}(n+1)}{n\omega
-x+c}.
\end{eqnarray}
Here, the superindices $+$ are omitted.

The $Z_2$ symmetry reflects into a symmetry in the eigenfunction:
$\varphi_2(-z)=C\varphi_1(z)$, where $C$ is an arbitrary  constant.
Without loss of generality we take $\varphi_{1,2}$  normalized
and real.  In this case,  $C=\pm1$, and  the transcendental function $G$ is
\begin{equation}
G^{\lambda }_{\pm}(x;z)=\varphi_2(-z)\mp \varphi_1(z)=0 \qquad \forall  \ z\in {\cal C}
\end{equation}
Setting $z=0$ as the above section, and substituting $\varphi
_{1,2}$ by $\phi _{1,2}$,
\begin{eqnarray}
G^{\lambda }_{+}(x)&=&-\xi\phi_1+\sqrt{\lambda}\phi_2, \nonumber \\
G^{\lambda }_{-}(x)&=&\sqrt{\lambda}\phi_1+\xi^* \phi_2.
\end{eqnarray}

\begin{figure}
\includegraphics[width=8cm]{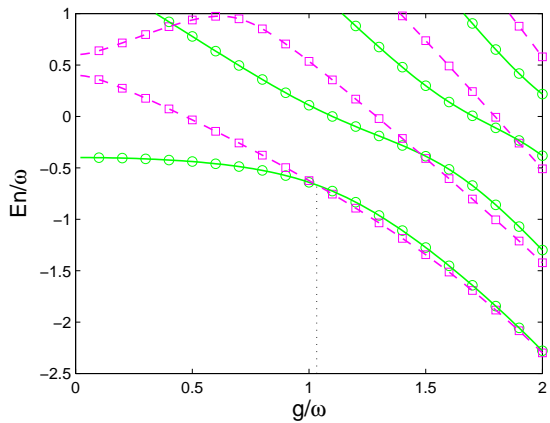}
\caption{(color online) Ground state level crossing.
The ground state energy and the first excited state energy
are crossed for anisotropic Rabi model with $Z_2$ symmetry.
The parity of the ground state changes when passing through
the crossing point. Here $\epsilon =0$ for keeping $Z_2$ symmetry,
we choose $\Delta =0.4, \lambda =0.5$,
so $g=\frac{4}{\sqrt{15}}, E_0^e=-\frac{2}{3}$ due to equations
(\ref{speciale},\ref{specialg}).
The green-solid lines are for positive parity, the purple-dashed lines
are for negative parity.
The circles and blocks are numerical data, they agree well
with analytical results shown in lines. Part of this figure
is shown in the upper panel of FIG.5. }\label{Figure6}
\end{figure}

The energy spectrum can be
divided into two cases.
One case is the regular solution which is solely determined by zeros of the transcendental function.
Another case corresponds to the exceptional solutions.
For this case, we can consider first the
the poles of the transcendental function
determined by setting $a_n(x)=0$,
\begin{eqnarray}
x_n^{pole}=n\omega-\frac{(1-\lambda)^2 g^2}{2\omega}.
\label{addpole}
\end{eqnarray}
At the same time, if  $K_{n+1}(x_n^{pole})=0$ for special
values of the parameters $g$ and $\Delta$,  the poles can be lifted,
because the numerator of $G_{\pm}^\lambda$ is also vanishing.
FIG. \ref{Figure5} shows the transition between regular solutions to
exceptional solutions.
This special solutions are Judd type solutions for the anisotropic Rabi model,
corresponding to the so called isolated integrability \cite{Judd}. Owing to
$G_{\pm}^\lambda \neq 0$, these eigenvalues have no definite parity,
and a double degeneracy of the eigenvalues occurs (see FIG. 3(b) in main text).
In this case,  the infinite series solutions $K_n$ and $L_n$ can be terminated as finite series
solutions.
We note that, in particular, there is no crossing with the same parity.
Incidentally, a crossing between the ground state and the first excited state
occurs for the anisotropic case which
corresponds to the exact solutions obtained by Judd \cite{Judd},
as already shown in Section IV.

FIG. \ref{Figure6} shows that the parity of the ground state changes sign when passing through
the level crossing point. When $g/\omega $ is small, the ground state is
positive parity, the first excited state is negative parity;
after passing through the crossing point where the parity changes sign,
the ground state is negative parity
and the first excited state is positive parity.
This parity changing can be demonstrated by the intrinsic symmetries of the ground state.

The ground state of Hamiltonian (\ref{ha}) with vanishing $\epsilon $ can be written as,
\begin{eqnarray}
\Psi (z)=\left( \begin{array}{c}
\varphi _1(z)\\ \varphi _2(z)
\end{array}
\right),
\end{eqnarray}
or $\varphi _1(z)\rightarrow \varphi _2(-z)$, $\varphi _2(z)\rightarrow \varphi _1(-z)$
since of the $Z_2$ symmetry. Additionally when $g/\omega $ is small which
is less than the crossing point value, the ground state is
parity positive and can be simplified as,
\begin{eqnarray}
\Psi _{+}(z)=\left( \begin{array}{c}
\varphi _1(z)\\ \varphi _1(-z)
\end{array}
\right).
\label{parity+}
\end{eqnarray}
In comparison, when $g/\omega $ is larger than the crossing point value, the
ground state is parity negative and takes the form
\begin{eqnarray}
\Psi _{-}(z)=\left( \begin{array}{c}
\varphi _1(z)\\ -\varphi _1(-z)
\end{array}
\right).
\label{parity-}
\end{eqnarray}
However, we find that both $\Psi _{\pm }(z)$ are not the eigenstates of the Hamiltonian at
the ground state level crossing point.

We may reformulate the ground state with different parities as
\begin{eqnarray}
\Psi _{+}(z)=\left( \begin{array}{c}
\varphi _1(z)+\varphi _2(-z)\\ \varphi _2(z)+\varphi _1(-z)
\end{array}
\right),
\label{parity1+}
\\
\Psi _{-}(z)=\left( \begin{array}{c}
\varphi _1(z)-\varphi _2(-z)\\ \varphi _2(z)-\varphi _1(-z)
\end{array}
\right).
\label{parity1-}
\end{eqnarray}
Those two states are the ground state and the first excited state.
They are the correct eigenstates corresponding to different parities
in the whole region including the level crossing point.
Explicitly, one may confirm that $\Psi _{\pm }(z)$ in (\ref{parity1+},\ref{parity1-})
are similar as the results in (\ref{parity+},\ref{parity-}) when $g/\omega $
is not at the crossing point, respectively.
The ground energy degeneracy at the crossing point also implies that
arbitrary linear combinations of states $\Psi _{+}(z)$ and $\Psi _{-}(z)$ in (\ref{parity1+},\ref{parity1-})
are also the ground state eigenstates.

With the help of the solutions (\ref{solution1},\ref{solution2}), and also
considering the transformation (\ref{W-transform}), the eigenstates at
the level crossing point can be written as follows, up to a whole factor,
\begin{eqnarray}
W\Psi _{+}(z)=\left( \begin{array}{c}
\sqrt{\lambda }(\exp(-\frac {\sqrt {\lambda }g\xi }{\omega }z)-\exp(\frac {\sqrt {\lambda }g\xi }{\omega }z))
\\
\xi ^{\dagger }(\exp(-\frac {\sqrt {\lambda }g\xi }{\omega }z)+\exp(\frac {\sqrt {\lambda }g\xi }{\omega }z))
\end{array}
\right),\\
W\Psi _{+}(z)=\left( \begin{array}{c}
\sqrt{\lambda }(\exp(-\frac {\sqrt {\lambda }g\xi }{\omega }z)+\exp(\frac {\sqrt {\lambda }g\xi }{\omega }z))
\\
\xi ^{\dagger }(\exp(-\frac {\sqrt {\lambda }g\xi }{\omega }z)-\exp(\frac {\sqrt {\lambda }g\xi }{\omega }z))
\end{array}
\right).
\end{eqnarray}
We remark that because of the ground states energy degeneracy, the linear combinations of the
ground states may lead to simpler solutions,
\begin{eqnarray}
\Psi (z)=\left( \begin{array}{c}
\sqrt {\lambda }\exp(-\frac {\sqrt {\lambda }g}{\omega }z)
\\
\exp(-\frac {\sqrt {\lambda }g}{\omega }z)
\end{array}
\right);
~
\left( \begin{array}{c}
\sqrt {\lambda }\exp(\frac {\sqrt {\lambda }g}{\omega }z)
\\
-\exp(\frac {\sqrt {\lambda }g}{\omega }z)
\end{array}
\right).
\end{eqnarray}
It can be checked that these are eigenstates of the
Hamiltonian (\ref{ha}) for vanishing $\epsilon =0$ and $\theta =0$,
with degeneracy conditions (\ref{speciale},\ref{specialg}).

Finally, we remark that the Bargmann representation of bosonic operator used in
our model can also be used in the JC mdoel, which has $U(1)$
symmetry. The eigenvalues and eigenstates can be obtained in the
familiar steps, but simpler because
the total number $\hat{N}=a^\dagger a+\frac{1}{2}(\sigma_z+1)$ is
conserved.

\subsection{Fitting with experimental data for the superconducting circuits in the strongly coupled regime}
\label{details_superconducting}
We consider the energy gaps between the anisotropic
Rabi model and the JC model. Such energy differences, generalizing the Bloch-Siegert effect of  the isotropic Rabi model, play important roles in many physical applications where the strong coupling regime is the actual  one,
like in the superconducting circuits and some similar physical systems \cite{Mooij1999,Gunter,Wallraff,Schuster,Peropadre,Nakamura}.

Ordinarily, there is no general form
for the gap, but we can analyze it at the degenerate points.
The ground state energy of the JC model is $E_0^{JC}=-\Delta $, so
the ground state gap at $|\Delta|=(1-\lambda^2)g^2/2\omega$ is,
\begin{eqnarray}
\Delta E_0=-\Delta -E_0^e= \frac {\lambda ^2g^2}{\omega }.
\end{eqnarray}
when $\lambda=0$ (the JC limit), the gap vanishes; for $\lambda=1$, the gap is just the standard
Bloch-Siegert shift in the Rabi model, $\Delta E_0={g^2}/{\omega }$. 
For $\lambda \neq 1$, the first excited state
energy gap with the  JC is
\begin{eqnarray}
\Delta E_1= \frac{\omega}{2} -\sqrt{(\Delta-\frac{\omega}{2})^2+g^2}
+\frac{(1+\lambda^2)g^2}{2\omega}.
\end{eqnarray}
For small $g/\omega$,  $\Delta E_1\approx \lambda
^2g^4/\omega^3$, remarkably  different from the standard Bloch-Siegert
shift ${g^2}/{\omega }$.
This is for case $|\Delta|=(1-\lambda^2)g^2/2\omega$, as we just mentioned.

In the Rabi model, there is a crossing between the second and the
third energy levels at $|\Delta|=\sqrt{\omega^2-4g^2}$,
$E=\omega-g^2/\omega$ \cite{Braak}, the second energy level of the
JC model in small $g/\omega$ can be written as
\begin{eqnarray}
E_2^{JC}&=&\frac{\omega}{2}+\sqrt{(\Delta-\frac{\omega}{2})^2+g^2}
\nonumber \\
&\approx &\omega-\frac{g^2}{\omega} +\frac{g^4}{\omega^3}.
\end{eqnarray}
Obviously, the second excited energy difference between the JC model
and Rabi model is in $g^4/\omega^3$ scale, too. However, we compare
the third excited energy difference at this point as
\begin{eqnarray}
E_3^{JC}
-E_1^{e,Rabi}&=&\frac{3\omega}{2}-\sqrt{(\Delta-\frac{\omega}{2})^2+2g^2}-
(\omega-\frac{g^2}{\omega})
\nonumber \\
&\approx &\frac{g^2}{\omega}
-\frac{2g^4}{\omega^3}.
\end{eqnarray}
where the condition $g/\omega \ll 1$ is used, the difference is
still in $g^2/\omega$ scale. Maybe the energy differences of the
second and third excited state are strange at the degenerate point.
Those differences may be verified by the recent experiments with
ultrastrong coupling.

In ultrastrong coupling regime, the deviation from the JC model
known as Bloch-Siegert shift was experimentally observed\cite{Mooij},
which is an LC resonator magnetically coupled to a
superconducting flux qubit in the ultrastrong coupling regime, and
the system can be modeled by the Hamiltonian,
\begin{eqnarray}
H'&=&\frac{ \omega_q}{2}(\sigma_z+1) + \omega_r a^\dagger a
\nonumber \\
&&+g(\cos\vartheta \sigma_z-\sin\vartheta \sigma_x)(a+a^\dagger),
\label{Mooij-Hamil}
\end{eqnarray}
with $\omega_q\equiv \sqrt{\epsilon^2+\Delta^2}$, $\epsilon=2\pi
I_p(\Phi-\Phi_0/2)$ and $\tan\vartheta=\Delta/\epsilon$, where
$\hbar$ is conventionally set to 1.

Following results shown in Ref.\cite{Mooij},
if we neglect the term $g\cos\vartheta \sigma_z(a+a^\dagger)$, which
only contributes a constant $-g^2\cos^2 \vartheta/\omega_r$ to the second
order under
the transformation $U=\exp(-g\cos\vartheta/\omega_r\sigma _z(a-a^{\dagger }))$.
And by omitting the counter-rotating term, the corresponding JC
model is given by
\begin{eqnarray}
H_{JC}=\frac{\omega_q}{2}(\sigma_z+1) +\omega_r a^\dagger a - g
\sin\vartheta (\sigma^- a^\dagger +\sigma^+ a).
\nonumber \\
\label{Mooij-Hamil-JC}
\end{eqnarray}
In the ultrastrong coupling regime, the rotating wave approximation
is thus inappropriate, the experimental results of
this system will not agree with the JC model. The Bloch-Siegert shift
caused by the counter-rotating term is evidently observed.

Then, we directly use our proposed anisotropic model to fit this
system
\begin{eqnarray}
H_{a-Rabi}&=&\frac{ \omega_q}{2}(\sigma_z+1) +\omega_ra^\dagger a
\nonumber \\
&&- g \sin\vartheta (\sigma^- a^\dagger +\sigma^+
a+\lambda (\sigma^- a +\sigma^+ a^{\dagger })).
\nonumber \\
\label{sup-gRJC}
\end{eqnarray}
where the anisotropic parameter $\lambda$ is decided by fitting, and
$g=0.74$GHz, $\Delta =4.21$GHz, $I_p=500$nA, $\omega _r=8.13$GHz,
are the same as previous obtained \cite{Mooij}. As shown in Fig. 4,
we can find that the experimental data agree
perfectly with the case $\lambda =0.5$ (red dashed dot line) which
is neither the Rabi model nor the JC model, and we remark that the
result of $\lambda =0.7$ case seems similar as the Hamiltonian
(\ref{Mooij-Hamil}) (solid black line), and $\lambda =0$ case is the
same as the JC Hamiltonian (\ref{Mooij-Hamil-JC}) (dashed black
line). Here we try to comment that by this experimental set up, the
anisotropic Rabi model may be tested in a full regime by using
qubit devices with strength of coupling ranging from weak to ultrastrong
up to $g\approx 2$GHz within current technologies.

\end{appendix}

\end{document}